# Crystal growth and characterization of Fe$_{1+\delta}$Se$_{1-x}$Te$_x$ (0.5 ≤ x ≤ 1) from LiCl/KCl flux

Qiaoyu Wang[a,c], Kexin Bi[a,d], Lewei Chen[a,b], Yunqing Shi[a,b], Junkun Yi[a,b], Yadong Gu[a], Menghu Zhou[a], Binbin Ruan[a], Xingye Lu[c], Mingwei Ma[a,*], Genfu Chen[a,b], Zhian Ren[a,b]

[a]Beijing National Laboratory for Condensed Matter Physics, Institute of Physics, Chinese Academy of Sciences, Beijing, 100190, China
[b]University of Chinese Academy of Sciences, Beijing, 100049, China
[c]Center for Advanced Quantum Studies and Department of Physics, Beijing Normal University, Beijing, 100875, China
[d]School of Physics and Microelectronics, Zhengzhou University, Zhengzhou 450052, China
*Email: mw_ma@iphy.ac.cn

**Abstract**

An eutectic LiCl/KCl flux method in a horizontal configuration has been used to grow a series of homogeneous Fe$_{1+\delta}$Se$_{1-x}$Te$_x$ single crystals of high quality with 0.5 ≤ $x$ ≤ 1. Compared with previously used melt-growth method, the stable crystallization process in LiCl/KCl flux below their peritectic temperatures results in better homogeneity and crystalline perfection identified by energy dispersive spectrometer and x-ray diffraction. The interstitial Fe value $\delta$ remains small within 0.5 ≤ $x$ ≤ 0.85 where the superconducting temperature $T_C$ is not sensitive to the Te content with sharp superconducting transition widths $\Delta T_C$ < 1 K and a maximum of $T_C$ = 14.3 K at $x$ = 0.61. The value $\delta$ starts to increase quickly accompanied by a deviation of linear behavior of crystal lattice parameters as well as the broadening of $\Delta T_C$ = 2.1 K at $x$ = 0.91, then suddenly rises up to $\delta$ > 0.1 followed by the disappearance of superconductivity and emergence of antiferromagnetic order at $x$ ≥ 0.96. We also observed a metallic to semiconducting transition in the normal state resistivity of Fe$_{1+\delta}$Se$_{1-x}$Te$_x$ with increasing Te content which is related to a localized electronic state induced by the interstitial Fe. The interstitial Fe value $\delta$ might be a key physical parameter to understand various properties of Fe$_{1+\delta}$Se$_{1-x}$Te$_x$ system.

## 1. Introduction

The Fe$_{1+\delta}$Se$_{1-x}$Te$_x$ system is an ideal platform for investigating the mechanism of iron-based superconductivity and is also promising in practical applications due to its simple crystal structure and unique properties [1-6]. Although intensive studies have been made on the properties of bulk Fe$_{1+\delta}$Se$_{1-x}$Te$_x$ materials for more than 15 years, some questions or controversies still remain regarding the crystal quality and the resulting physical properties [7-15]. The topological nature of Fe$_{1+\delta}$Se$_{1-x}$Te$_x$ causes intense debate. First-principles band-structure calculations for $Fe_{1+\delta}Se_{1-x}Te_x$ revealed that topological nature indeed emerges upon Te substitution [16-18]. In the higher Te-substituted $FeSe_{0.45}Te_{0.55}$, the spin-polarized Dirac surface state was observed from ultrahigh resolution laser-based spin angle-resolved photoemission spectroscopy (ARPES) [19] and the scanning tunneling spectroscopy experiment revealed signatures of Majorana bound state (MBS) [20]. However, a subsequent experiment performed under similar conditions observed that the bound-state peaks appear only at finite energies, suggesting their trivial nature [21]. This is in particular observed in materials produced by melting routes, where chemical inhomogeneities and multiple phases are commonly obtained [7-15]. In principle, high-quality $Fe_{1+\delta}Se_{1-x}Te_x$ single crystals cannot be directly grown from their own melt as they are incongruent melting compounds with varying peritectic temperatures of FeSe (457 ℃) and FeTe (847 ℃) according to their binary phase diagrams [22-23]. Even no complete ternary phase diagram has been assessed for the mixed compound $Fe_{1+\delta}Se_{1-x}Te_x$, synthesis of $FeSe_{0.5}Te_{0.5}$ materials via melting routes leads inevitably to the presence of Se-rich secondary phases as dendritic precipitates in a tetragonal matrix [10-12]. Also, the presence of excessive amounts of Fe further complicates the investigation of the properties of $Fe_{1+\delta}Se_{1-x}Te_x$, spanning from its superconducting to normal state characteristics [24-30]. Therefore, a more suitable growth route is in urgent need to obtain more homogeneous $Fe_{1+\delta}Se_{1-x}Te_x$ single crystals of high quality which is the prerequisite for studying their intrinsic properties including the correspondence between the zero-energy conductance peak in the vortex core of $Fe_{1+\delta}Se_{1-x}Te_x$ and the MBS [3].

Usually, high-temperature solution method is the most suitable way for growing incongruent melting compounds, which can result in better crystal quality with respect to point defects, dislocation densities and composition distribution compared to crystals grown directly from their own melt. The key to this issue is how to choose the suitable flux with low melting point, high solubility and no contamination for $Fe_{1+\delta}Se_{1-x}Te_x$. Recently, the $KCl/AlCl_3$ molten salt method with growth temperature below their peritectic temperature was successfully employed to grow homogeneous and phase-pure $Fe_{1+\delta}Se_{1-x}Te_x$ single crystals with lower Te substitution range ($x \leq 0.5$) [31-32]. The conditions of the crystal growth are essential not only for the quality of the crystals but also for the Te substitution level $x$ of FeSe. The $KCl/AlCl_3$ flux is also suitable for S/Co/Cu substituted FeSe single crystals [33-37] but cannot grow $Fe_{1+\delta}Se_{1-x}Te_x$ single crystals with higher Te substitution ($x \geq 0.5$) [31-32]. In this paper, we opted for an alternative approach using LiCl/KCl molten salt (or flux) to grow a series of $Fe_{1+\delta}Se_{1-x}Te_x$ single crystals with improved homogeneity and crystalline perfection. With $x$ continuously increasing from 0.5 to 1, we have observed a transition from metallic to semiconducting behavior in the normal state resistivity of $Fe_{1+\delta}Se_{1-x}Te_x$. When the Te content exceeded a certain threshold $x \geq 0.96$, superconductivity disappears and antiferromagnetic order emerges in $Fe_{1+\delta}Se_{1-x}Te_x$ accompanied by a sudden increase of interstitial Fe and a deviation from linear behavior of the crystal lattice parameters.

## 2. Experimental methods

High purity Fe, Se, Te powders are ground and mixed with an agate mortar and pestle in glove box. The nominal composition $x_{raw}$ for each batch was Fe : Se : Te = 1.1 : (1-$x_{raw}$) : $x_{raw}$ in molar ratio, with the mole number $x_{raw}$ varying from 0.5 to 1 as shown in Table 1. The mixtures were then pressed into separate columnar shapes and sealed in an evacuated quartz tube. The quartz tube was placed in a box furnace and heated at 700 ℃ for 10 hours. The obtained products were once again well ground, sealed in evacuated quartz tube and heated at 700 ℃ for 10 hours. After that, the materials were ground into fine powders as precursor in the glove box. The $Fe_{1+\delta}Se_{1-x}Te_x$ solid solutions (~ 1 g) are mixed with LiCl/KCl salt (~ 9 g) with LiCl : KCl = 3 : 2 in molar ratio, which has the eutectic point at ~ 350 ℃. The mixtures with total mass ~ 10 g are transferred into a long $Al_2O_3$ crucible (Φ18 mm × 300 mm) and the crucible is placed into a long quartz tube (Φ30 mm × 400 mm). The mouth of crucible was covered with a lid and sealed by a high temperature glue as shown in Fig. 1(a) in order to avoid the leakage of flux from crucible at high temperature. The quartz tube was then sealed and placed in a horizontal double zone furnace. The double zones were heated to 600℃ (hotter zone) and 500℃ (cooler zone) resulting in a stable temperature gradient 4 ℃/cm as displayed in Fig. 1(a). After a growth duration of 20 days, $Fe_{1+\delta}Se_{1-x}Te_x$ single crystals with plate-like forms immersed in LiCl/KCl salt were obtained near the cooler portion of the crucible as shown in Fig. 1(b). The single crystals were extracted by dissolving the LiCl/KCl flux in a beaker filled with distilled water. The single crystals were then removed from water with tweezers. Dozens of pieces of $Fe_{1+\delta}Se_{1-x}Te_x$ single crystals are obtained from each batch with typical lateral sizes up to 1－2 mm as displayed in Fig. 1(c).

In order to more accurately determine the tiny Fe atoms at the interstitial sites, the chemical composition of as-grown $Fe_{1+\delta}Se_{1-x}Te_x$ single crystals was determined by inductively coupled plasma atomic emission spectroscopy (ICP), where we define $x_{ICP}$ as the real concentration of Te and $\delta$ value as the excess Fe content as listed in Table 1. The micro-morphology of $Fe_{1+\delta}Se_{1-x}Te_x$ single crystals was examined by scanning electron microscope (SEM) on Phenom ProX. The composition homogeneity of the surface of the crystals was checked by energy dispersive spectrometer (EDS) on Phenom ProX. The Laue picture was taken on the surface of crystal plate by the x-ray back-Laue photography. Powder and single crystal x-ray diffraction (XRD) measurements for phase identification of $Fe_{1+\delta}Se_{1-x}Te_x$ were carried out at room temperature on an x-ray diffractometer (Rigaku UltimaIV) using Cu $K_\alpha$ radiation. The crystal lattice parameters are refined by the Rietveld Analysis method using the Highscore software. The resistivity of $Fe_{1+\delta}Se_{1-x}Te_x$ single crystals was measured on Quantum Design PPMS-9 using the standard 4-probe method with silver paste for contact. The magnetic measurements were carried out on a SQUID magnetometer (Quantum Design MPMS XL-1).

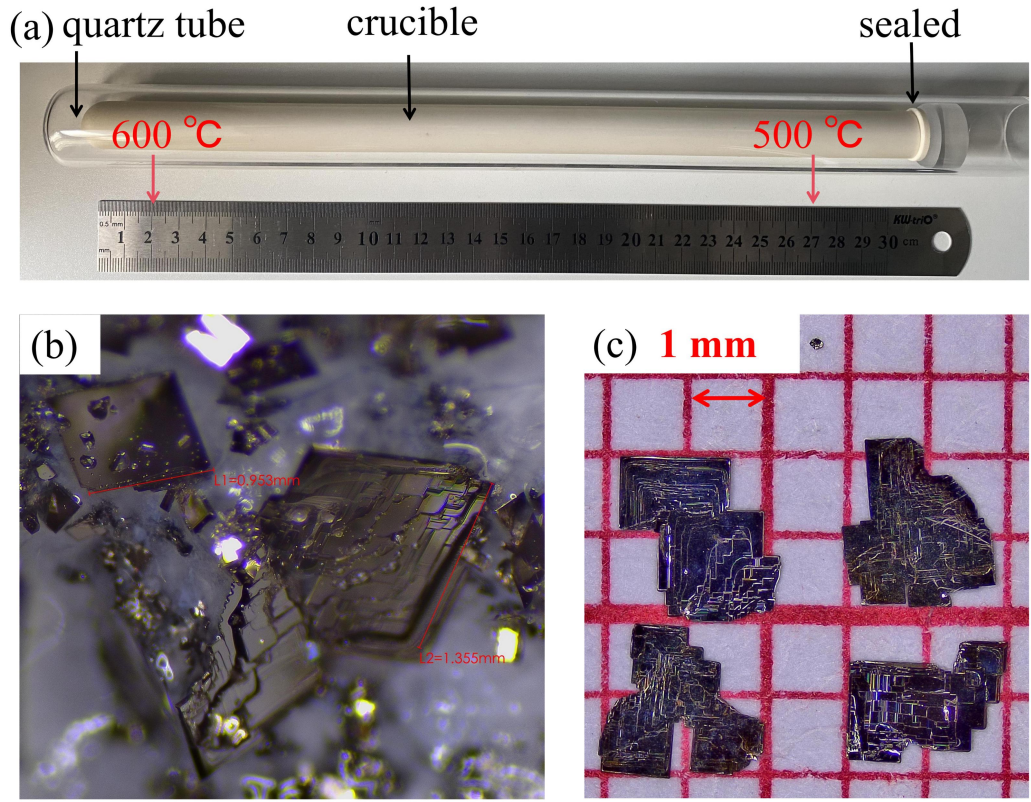

Fig. 1 (a) Schematic diagram for growth of $Fe_{1+\delta}Se_{1-x}Te_x$ single crystals. (b) Photograph of plate-like as-grown $Fe_{1+\delta}Se_{1-x}Te_x$ single crystals with shining surfaces immersed in salt at the cooler portion of crucible. (c) Typical picture of several selected $Fe_{1+\delta}Se_{1-x}Te_x$ single crystals with the lateral size of 1－2 mm.

## 3. Results and discussion

The inset of Fig. 2(a) is the back-scattering x-ray Laue patterns obtained on a cleavage surface of $Fe_{1+\delta}Se_{1-x}Te_x$ single crystal. It exhibits sharp and uniform diffraction spots, demonstrating the crystalline perfection of our as-grown single crystals. The 4-fold axis symmetry due to the tetragonal structure indicates that the $c$-axis is perpendicular to the cleavage surface, namely (0 0 1) plane. Figure 2 (a-i) illustrates the micro-morphology of $Fe_{1+\delta}Se_{1-x}Te_x$ single crystals with $x_{ICP}$ values ranging from 0.5 to 1 taken by SEM on (0 0 1) plane of crystal plates. All the crystal plates exhibit flat surfaces, clear tetragonal cleavages as well as the growth steps marked by the red circles, indicating the crystal growth rate of (1 0 0)/(0 1 0) plane is much faster than that of (0 0 1) plane.

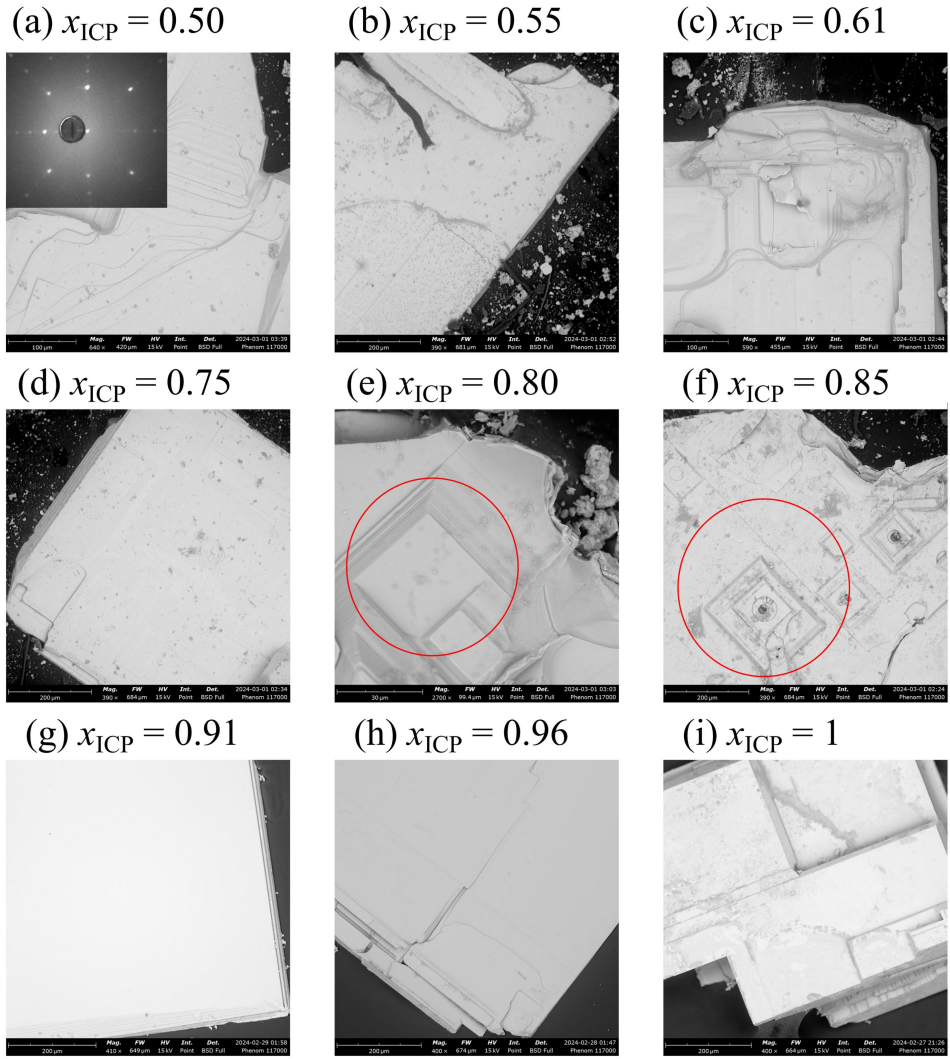

Fig. 2 (a-i) Micro-morphology of $Fe_{1+\delta}Se_{1-x}Te_x$ single crystals with $x_{ICP}$ values ranging from 0.5 to 1 taken by SEM on (0 0 1) plane of crystal plates where flat surface and tetragonal cleavage can be clearly seen and the growth steps are marked by the red circles. The inset of (a) is an x-ray back-Laue photograph on the cleavage surface of $Fe_{1+\delta}Se_{1-x}Te_x$ ($x_{ICP} = 0.5$) single crystal.

The chemical homogeneity of $Fe_{1+\delta}Se_{1-x}Te_x$ single crystals was examined by mapping the selected area of the (0 0 1) surface as shown in Fig. 3. The elements of Fe, Se and Te of as-grown crystals distribute homogeneously which are demonstrated by the yellow, purple and green mapping areas in contrast to the non-uniformity of as-grown crystals from melt-growth technique [7-15, 18-24]. Although the quality of as-grown samples from melt-growth technique can be improved after removing the excess Fe or reducing the impurity phase by various post-annealing treatments [15], the compounds of Fe in some form of oxides possibly remain in the crystal which might affect their intrinsic properties and the exact determination of the chemical composition by ICP and EDS. After identifying the homogeneity of our as-grown single crystals, their exact chemical composition are determined by ICP analysis as shown in Table 1 and Fig. 5(f). With increasing nominal composition $x_{raw}$, the real concentration of Te $x_{ICP}$ in $Fe_{1+\delta}Se_{1-x}Te_x$ single crystals has also been raised up, exhibiting a positive correlation between $x_{raw}$ and $x_{ICP}$, while the interstitial Fe value $\delta$ remains consistently small ($\delta < 0.05$) and shows a slight increase with increasing $x_{ICP}$ within $0.5 \leq x_{ICP} \leq 0.85$. The $\delta$ value starts to increase more quickly at $x_{ICP} = 0.91$ and rises up to $\delta > 0.1$ at $x_{ICP} \geq 0.96$. This implies that more interstitial Fe is advantageous for the structure stabilization of tetragonal $Fe_{1+\delta}Se_{1-x}Te_x$ in Te-rich area.

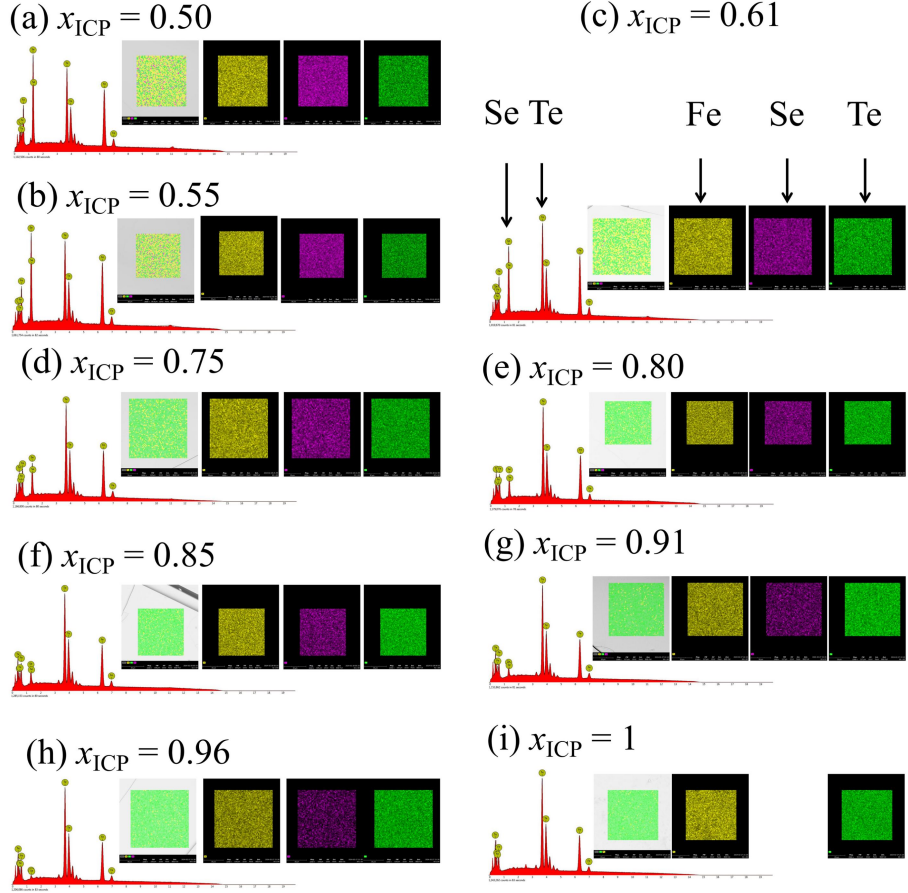

Fig. 3. (a-i) The chemical composition mapping of $Fe_{1+\delta}Se_{1-x}Te_x$ single crystals with $0.5 \leq x_{ICP} \leq 1$. The inset shows the distribution of Fe, Se and Te elements on the surface as indicated by the yellow, purple and green area.

Table 1. Growth conditions and physical parameters of $Fe_{1+\delta}Se_{1-x}Te_x$

| $x_{raw}$ | $x_{ICP}$ | $\delta$ | $a$ (Å) | $c$ (Å) | $T_C$ (K) | $\Delta T_C$ (K) | $T_N$ (K) |
|---|---|---|---|---|---|---|---|
| 0.5 | 0.50 | 0.031 | 3.7976 | 5.9887 | 13.2 | 0.9 | |
| 0.6 | 0.55 | 0.035 | 3.8008 | 6.0183 | 13.9 | 0.6 | |
| 0.65 | 0.61 | 0.038 | 3.8024 | 6.0662 | 14.3 | 0.6 | |
| 0.8 | 0.75 | 0.041 | 3.8102 | 6.1611 | 13.7 | 0.7 | |
| 0.85 | 0.80 | 0.045 | 3.8119 | 6.1941 | 12.9 | 0.5 | |
| 0.90 | 0.85 | 0.047 | 3.8174 | 6.2273 | 12.1 | 0.7 | |
| 0.95 | 0.91 | 0.061 | 3.8189 | 6.2503 | 11.6 | 2.1 | |
| 0.98 | 0.96 | 0.111 | 3.8194 | 6.2663 | | | 51 |
| 1 | 1 | 0.120 | 3.8204 | 6.2766 | | | 63 |

Powder and single crystal x-ray diffraction (XRD) patterns of $Fe_{1+\delta}Se_{1-x}Te_x$ are displayed in Fig. 4(a-b). All the diffraction peaks exhibit systematic shifts with the variation of Te content $x_{ICP}$. Only (0 0 1) reflections are observed from the single crystal XRD pattern indicating that the single crystals are in perfect (0 0 1) orientation as shown in Fig. 4(b). As illustrated in Fig. 4(a) the results of the powder XRD analysis indicate that all the diffraction peaks can be accurately indexed with a previously reported tetragonal structure, characterized by the space group of P4/nmm [1-2] and these samples show good quality without second phases. Within the range $0.5 \leq x_{ICP} \leq 0.85$, both the *a* and *c* lattice parameters as well as the cell volume *V* exhibit a linear increase with increasing Te substitution level $x_{ICP}$, as depicted in Fig. 5(a-c). This indicates that the successful substitution of Se by Te in $Fe_{1+\delta}Se_{1-x}Te_x$ crystal lattice as a result of the larger radius of Te atom than Se atom. However, when $x_{ICP}$ exceeds 0.91 we observe a deviation of linear behavior of crystal lattice parameters where we define the deviation rate as $\Delta = |V-V_{fitting}|/V_{fitting} \times 100\%$ as shown in Fig. 5(d). This deviation corresponds to (is associated with) a sudden increase in the value of $\delta$ at $x_{ICP} = 0.91$ as mentioned above in Table 1 and Fig. 5(f). This observation implies the close relationship between the $\delta$ value and the deviation of linear behavior of crystal lattices.

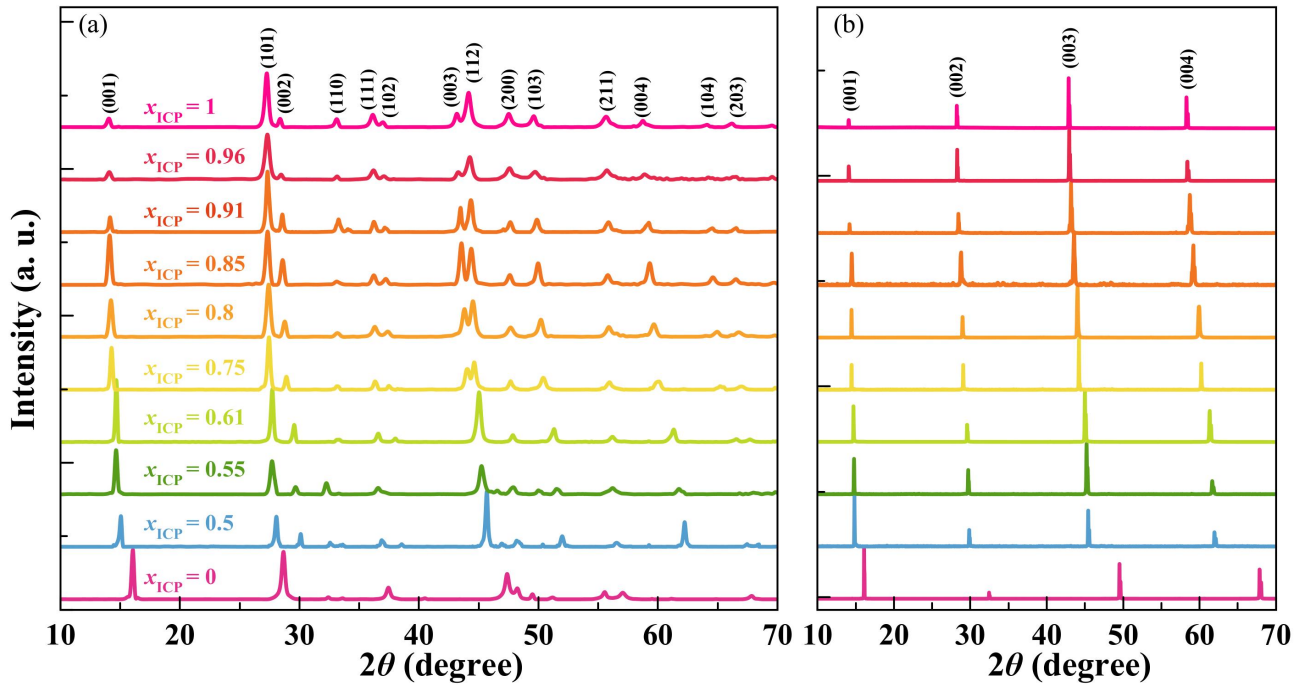

Fig. 4. (a) Powder and (b) single crystal XRD patterns of $Fe_{1+\delta}Se_{1-x}Te_x$ single crystals with $0.5 \leq x_{ICP} \leq 1$. The XRD data of FeSe grown by $AlCl_3$/KCl molten salt method are plotted here as reference.

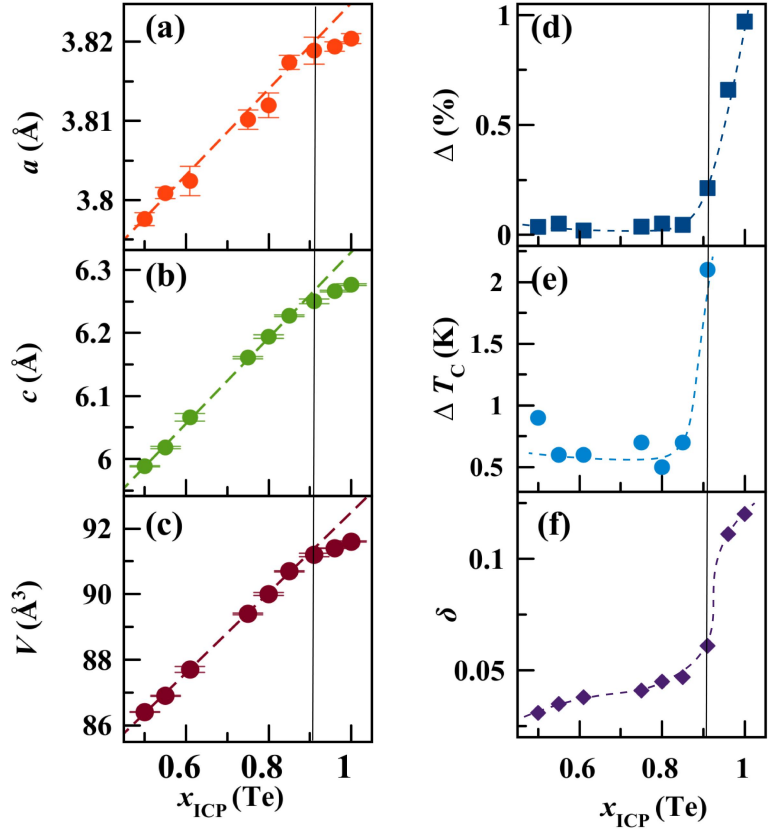

Fig. 5. Te substitution level $x_{ICP}$ dependence of (a) $a$-axis, (b) $c$-axis lattice parameters, (c) cell volume $V$, (d) cell volume deviation rate $\Delta$, (e) superconducting transition width and (f) interstitial Fe value $\delta$ of $Fe_{1+\delta}Se_{1-x}Te_x$ single crystals.

Figure 6 shows the temperature dependence of the normalized electrical resistivity ($\rho/\rho_{200\ K}$) for $Fe_{1+\delta}Se_{1-x}Te_x$. In the normal state ($T \geq T_C$ or $T_N$) as displayed in Fig. 6 (a-b), while the resistivity demonstrates a metallic behavior when Te content $x_{ICP} \leq 0.55$, a weak upturn with decreasing temperature appears around 58 K in the resistivity of $Fe_{1+\delta}Se_{1-x}Te_x$ at $x_{ICP} = 0.61$ indicating a metallic to semiconducting transition behavior as shown in the inset of Fig. 6(a). When Te content $x_{ICP} \geq 0.75$, the normal state resistivity increases gradually with cooling, showing a semiconducting behavior. For higher Te content with $x_{ICP} \geq 0.96$ as shown in Fig. 6(b), resistivity drops steeply around $T_N$ and then exhibits a metallic behavior at low temperatures. The discontinuous change is due to antiferromagnetic (AFM) transition with $T_N = 63$ K at $x_{ICP} = 1$ decreasing to $T_N = 51$ K at $x = 0.96$, which is also confirmed by the magnetization measurement in the inset of Fig. 6(b) in agreement with previous reports [24-30]. The normal state metallic to semiconducting transition is probably related to the excess Fe at interstitial site of $Fe_{1+\delta}Se_{1-x}Te_x$ which would result in a weakly localized electronic state [27]. The large amount interstitial Fe value $\delta > 0.1$ for Te content with $x_{ICP} \geq 0.96$ is responsible for their AFM order due to the enhancement of magnetic coupling between excess Fe and the adjacent Fe square sheets [27].

We turn to the influence of interstitial Fe on the superconductivity of $Fe_{1+\delta}Se_{1-x}Te_x$. Figure 6(c) shows the enlarged view of normalized resistivity from 2 K to 20 K. The $T_C$ defined here as the zero resistance temperature is around 10 − 15 K and not sensitive to the Te content within the range $0.5 \leq x_{ICP} \leq 0.91$ as displayed in Fig. 6(c). With increasing Te content $x_{ICP}$, the $T_C$ reaches a maximum of 14.3 K at $x_{ICP} = 0.61$, slightly decreases to $T_C = 11.6$ K at $x_{ICP} = 0.91$ and suddenly disappears at $x_{ICP} = 0.96$. The bulk superconductivity of as-grown crystals is confirmed by the magnetic susceptibility measurement with superconducting volume larger than 90% as shown in the inset of Fig. 6(c) in contrast to the non-superconducting samples grown from their own melt [24-30]. Here we define the superconducting transition width $\Delta T_C$ as the temperature range from 10% to 90% drop of resistivity at the onset temperature of superconductivity which are listed in Table 1 and shown in Fig. 5(e). The $\Delta T_C$ remains less than 1 K within $0.5 \leq x_{ICP} \leq 0.85$ indicating a sharp superconducting transition and becomes broad at $x_{ICP} = 0.91$ corresponding to the sudden increase of excess Fe value $\delta$ [Fig. 5(f)] as well as the deviation of linear behavior of crystal lattice in Fig. 5(d). Increased Fe content in the interstitial sites not only hinders the occurrence of superconductivity but also leads to the formation of a weakly localized electronic state. The magnetic interaction between the excess Fe and the nearby Fe square-planar sheets causes the disruption of the pairing of electrons necessary for superconductivity.

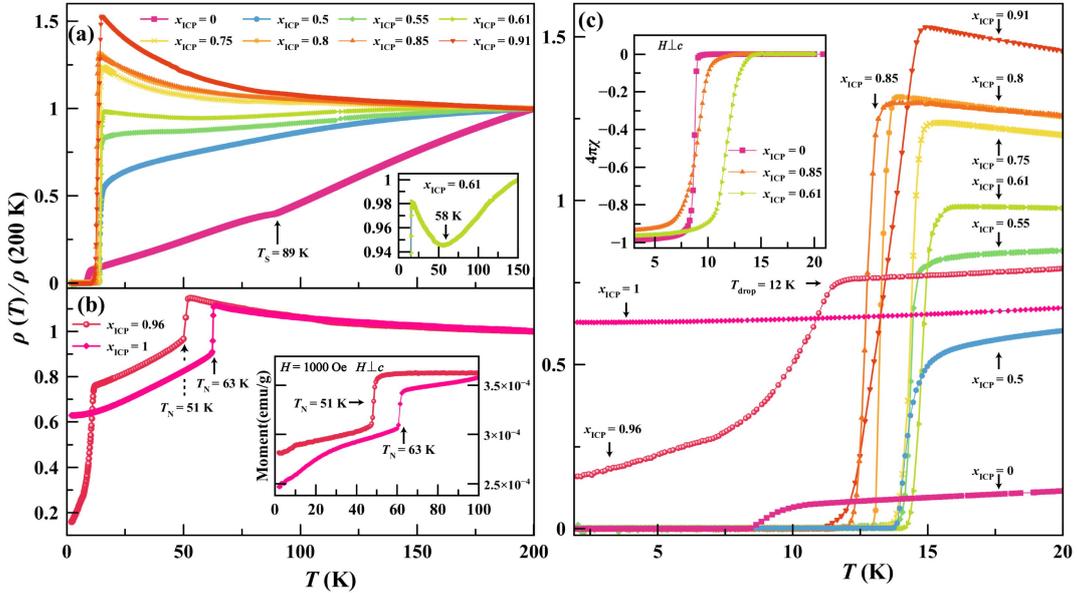

Fig. 6. (a-b) Temperature dependence of normalized resistivity of $Fe_{1+\delta}Se_{1-x}Te_x$ single crystals with $0.5 \leq x_{ICP} \leq 1$ from 2 K to 200 K. The resistivity data of FeSe are plotted here as reference. The inset in (a) shows the enlarged view of resistivity for $x_{ICP} = 0.61$, the inset in (b) shows the magnetization of $Fe_{1+\delta}Se_{1-x}Te_x$ for $x_{ICP} = 0.96$ and 1. (c) Enlarged view of normalized resistivity of $Fe_{1+\delta}Se_{1-x}Te_x$ single crystals with $0.5 \leq x_{ICP} \leq 1$ from 2 K to 20 K. The inset shows the magnetic susceptibility of $Fe_{1+\delta}Se_{1-x}Te_x$ for $x_{ICP} = 0$, 0.61 and 0.85.

## 4. Conclusion

We have found a suitable LiCl/KCl flux to grow a series of $Fe_{1+\delta}Se_{1-x}Te_x$ single crystals with the Te doping level ($0.5 \leq x \leq 1$). The as-grown single crystals are phase-pure and more uniform in contrast to the previous samples grown by melt-growth method. The lattice parameters of $a$ and $c$ increase with increasing Te content but deviate from the linear increase behavior with $x \geq 0.91$ which are related to the sudden increase interstitial Fe value $\delta$. Also, the interstitial Fe are responsible for the suppression of superconductivity and emergence of AFM order. The LiCl/KCl flux method opens a new route to obtain high-quality $Fe_{1+\delta}Se_{1-x}Te_x$ single crystals which are important for clarifying the key physical parameters with respect to superconductivity and also is applicable for growing other incongruent single crystals.

## 5. Acknowledgments

The work was supported by the National Key Research and Development of China (Grant No. 2022YFA1602800, 2018YFA0704200) and the National Natural Science Foundation of China (Grant No. 12004418) and the Strategic Priority Research Program of Chinese Academy of Sciences (Grant No. XDB25000000).